\documentclass[aps,prb,showpacs,floatfix,superscriptaddress,notitlepage]{revtex4-1}

\usepackage{amsmath,amsfonts,amssymb}
\usepackage{graphicx}
\usepackage[colorlinks=true, allcolors=blue]{hyperref}

\begin{document} 
\title{Relaxation of nonequilibrium populations after a pump: the breaking of Mathiessen's rule}

\author{J.~K.~Freericks}
\email{James.Freericks@georgetown.edu}
\affiliation{Department of Physics, Georgetown University,
              37th and O Sts. NW, Washington, DC~20057, USA}
\author{O.~Abdurazakov}
\affiliation{Department of Physics, North Carolina State University, Raleigh, NC~27695, USA}
\author{A.~F.~Kemper}
\affiliation{Department of Physics, North Carolina State University, Raleigh, NC~27695, USA}


\pagestyle{empty} 
\setcounter{page}{301} 

\begin{abstract}
From the early days of many-body physics, it was realized that the self-energy governs the relaxation or lifetime of the retarded Green's function. So it seems reasonable to directly extend those results into the nonequilibrium domain. But experiments and calculations of the response of quantum materials to a pump show that the relationship between the relaxation and the self-energy only holds in special cases. Experimentally, the decay time for a population to relax back to equilibrium and the linewidth measured in a linear-response angle-resolved photoemission spectroscopy differ by large amounts. Theoretically, aside from the weak-coupling regime where the relationship holds, one also finds deviations and additionally one sees violations of Mathiessen's rule. In this work, we examine whether looking at an effective transport relaxation time helps to analyze the decay times of excited populations as they relax back to equilibrium. We conclude that it may do a little better, but it has a fitting parameter for the overall scale which must be determined.
\end{abstract}


\maketitle


\section{Introduction}
\label{intro}

Back when many-body physics and Green's function methods were being established, it was noticed that the self-energy provides the lifetime for the equilibrium Green's function~\cite{galitski_migdal}. In particular, it is the imaginary part of the self-energy evaluated at the pole of the Green's function that lies closest to the real axis in the lower half plane that determines this relaxation rate. This result leads one to infer that for small deviations from equilibrium, the imaginary part of the self energy should continue to provide the relaxation rate for the electrons. But trying to make this work immediately leads to a complication. Namely, in equilibrium there is only one time, while in nonequilibrium there are two times associated with the Green's function. The decay rate determined from the linear-response analysis governs decay in the relative time direction, but the change in the population of the electrons in a given momentum state is governed by decay in the average time direction. While those decay rates might be related to one another, it is by no means obvious that they must be the same, and indeed, we often find they are not.

Hence, there is a need to re-evaluate how relaxation occurs in nonequilibrium, since it is not governed by the same behavior that drives lifetimes for equilibrium systems.

One thing that is always relevant for relaxation is the constraints provided by the phase space for scattering and the Pauli exclusion principle. This leads to the so-called ``phonon window effect,''~\cite{michael_prx,lex_prb} where scattering by an optical phonon is sharply reduced as one gets to energies that lie below the phonon frequency. In particular, the relaxation outside the phonon window $\omega>\Omega$ is quite rapid, but within the window $\omega<\Omega$, it becomes slow. As the system is excited, the Pauli blocking is reduced within the window and enhanced outside the window, so the relaxation rates move closer to each other, but, speaking quantitatively, they rarely become too similar.

These effects have already been seen in experiment. A direct comparison of the relaxation time for a population and the lifetime, as measured in linear-response angle-resolved photoemission spectroscopy (ARPES), showed large differences, sometimes more than an order of magnitude difference in the relaxation rates~\cite{patrick}. The phonon window effect and its change with pump fluence have also been observed~\cite{rameau}. Theory has started to examine these effects too. In particular, an equation of motion technique was used to determine the initial contributions to the relaxation rate at long times, and comparison with the imaginary part of the self-energy also showed significant differences, but not yet as large as seen in experiment~\cite{entropy,fortschritte}.

In this work, we approach the problem from a different perspective. It is well known from linear-response theory, that the relaxation of the current comes from the transport relaxation time, which is related to the imaginary part of the self-energy, but is clearly different as well. Here, we compute the generalization of that relaxation time to see whether it gives a relaxation rate that is closer to the relaxation rate that can be extracted directly from the time dependence of the populations. While we find some improvement, it is not significant, and it requires us to adjust an overall scaling factor for the rate, so it is not a complete determination of the relaxation rate. These results are intimately related to a breakdown of Mathiessen's rule. That rule says that when we have multiple scattering mechanisms, the relaxation rates for each mechanism add together to create a net relaxation rate. But in nonequilibrium relaxation, we often see multiple rates arise in different time ranges due to different relaxation processes and bottlenecks to energy transfer; we do not go into full detail of that phenomenon here, as we instead focus on trying to identify the primary relaxation mechanism for the electron-phonon interaction.

We work with the Holstein model, which involves a single band of uncorrelated electrons that interact with an optical phonon (Einstein mode) via a density-coordinate coupling. The Hamiltonian is
\begin{equation}
\mathcal{H}=-\sum_{ ij\sigma}t_{ij}c^\dagger_{i\sigma}c^{\phantom{\dagger}}_{j\sigma}-
\mu\sum_{i\sigma}c^\dagger_{i\sigma}c^{\phantom{\dagger}}_{i\sigma}-
g\sum_i c^\dagger_{i\sigma}c^{\phantom{\dagger}}_{i\sigma}\left (b_i^\dagger+b^{\phantom{\dagger}}_i\right )+\Omega\sum_i b^\dagger_ib^{\phantom{\dagger}}_i
\label{eq: ham}
\end{equation}
where $c_{i\sigma}^{\phantom{\dagger}}$ ($c_{i\sigma}^\dagger$) destroys (creates) an electron with spin $\sigma$ at lattice site $i$, $b_i^{\phantom{\dagger}}$ ($b^\dagger_i$) are the phonon lowering  (raising) operators for the optical phonon at site $i$, $-t_{ij}$ is the hopping integral that connects site $i$ with site $j$,  $\mu$ is the electron chemical potential, $g$ is the electron-phonon coupling, and $\Omega$ is the phonon frequency. The dimensionless electron-phonon coupling $\lambda$ is given by the slope of the real part of the self-energy in equilibrium; here we have $g^2=0.02$~eV, $\Omega=0.1$~eV and $\lambda\approx 0.34$. We work on a square lattice with only a nearest neighbor hopping of $0.25$~eV. The system is at half-filling, which results when we choose $\mu=0$ and ignore the Hartree term in the perturbation theory (which we do because the Hartree term only shifts the chemical potential). The initial temperature is chosen to be $T=0.025$~eV (room temperature). The electric field is described in the Hamiltonian gauge, where ${\bf E}(t)=-d{\bf A}(t)/dt$, where we set $\hbar=c=1$. Then, we use the Peierls' substitution in the bandstructure to incorporate the field into the Hamiltonian. If we write
\begin{equation}
\epsilon({\bf k})=-\sum_j t_{ij}e^{i{\bf k}\cdot {\bf R}_j}-\mu
\label{eq: bandstructure}
\end{equation}
with ${\bf k}$ the momentum and ${\bf R}_j$ the position vector for the lattice site $j$. Then the Peierls substitution is $\epsilon({\bf k})\rightarrow \epsilon({\bf k}-{\bf A}(t))$. We start the system off in equilibrium at a temperature $T$, and then turn on an electric field, which we assume to be spatially uniform. We ignore all magnetic field effects. The vector potential is a sinusoidal oscillating wave with a Gaussian envelope. The amplitude of the vector potential is 0.5, the standard deviation of the Gaussian is 10~eV$^{-1}$ and the oscillation frequency of the sinusoidal wave is 0.5~eV. This wave has 4-5 visible periods and is visibly nonzero over a total range of about 50~eV$^{-1}$.

\begin{figure}
 \centerline{\includegraphics[height=0.1\textheight]{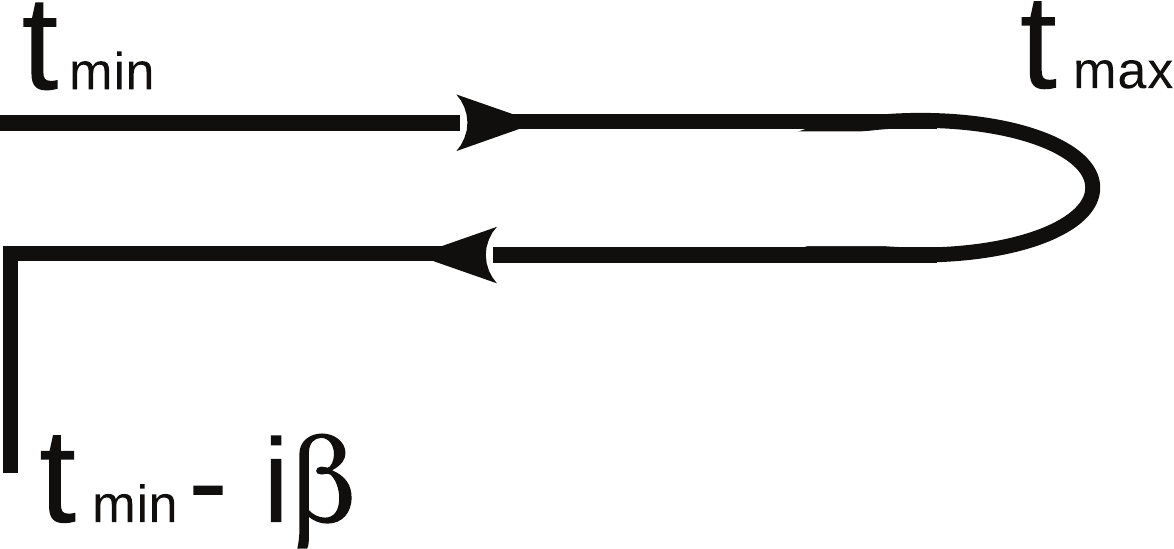}}
 \caption{Kadanoff-Baym-Keldysh time contour, which runs from a minimum time to a maximum time along the real time axis, then backwards to the minimum time, and then 
 parallel to the imaginary axis for a length given by the inverse of the initial equilibrium temperature.}
 \label{fig: contour}
\end{figure}

The contour-ordered Green's functions depend on two times, each lying on the Kadanoff-Baym-Keldysh contour shown in Fig.~\ref{fig: contour}. The local Green's function is defined via
\begin{equation}
G^c_{ij\sigma}(t,t')=-i\langle\mathcal{T}_c c^{\phantom{\dagger}}_{i\sigma}(t)c^\dagger_{j\sigma}(t')\rangle
\label{eq: green}
\end{equation}
where the angle brackets denote an average with respect to the initial equilibrium distribution
\begin{equation}
\langle \mathcal{O}(t)\rangle={\rm Tr} e^{-\beta\mathcal{H}(t\rightarrow-\infty)}\mathcal{O}(t)\frac{1}{\mathcal{Z}},
\label{eq: ave}
\end{equation}
$\beta=1/T$ is the inverse temperature of the initial equilibrium distribution, and $\mathcal{Z}={\rm Tr}\exp[-\beta\mathcal{H}(t\rightarrow-\infty)]$ is the partition function for the initial equilibrium state. The time-dependence of the operators is expressed in the Heisenberg picture. The symbol $\mathcal{T}_c$ is the time-ordering operator on the contour, which places later objects, according to where they sit on the contour, to the left.

We employ Migdal-Eliashberg theory as the impurity solver for the nonequilibrium version of dynamical mean-field theory (DMFT)~\cite{frturzlat_prl97,fr_prb77} to solve this problem by employing the Kadanoff-Baym-Keldysh formalism~\cite{kadanoff,keldysh}. The strategy is briefly summarized in the next section, where we discuss the perturbation theory and the techniques employed to solve Dyson's equation.
We also show the explicit formulas used for the data analysis.

\section{Formalism}
\label{sec:1}

Migdal-Eliashberg theory is employed to determine the local self-energy because the phonon energy scale is much smaller than the electron energy scale, implying we can neglect vertex corrections. Instead of how Migdal-Eliashberg theory is employed for linear-response and equilibrium calculations, where the phonons are the dressed phonons, so they are not renormalized, we self-consistently dress the phonons for the Holstein model by solving for the electron and phonon self-energies from a common conserving approximation. We simultaneously perform a self-consistent perturbation theory for the electronic Green's function; this approach allows us to take into account the finite heat capacity of the phonons and have their properties transiently change as they absorb energy. Since Migdal-Eliashberg theory involves a local self-energy, it is a form of DMFT, and we employ the NEDMFT approach to solving the problem. Note that all objects are contour-ordered continuous matrix operators, which depend on two times, each lying on the Kadanoff-Baym-Keldysh contour.

We start with a guess for the self-energy and then determine the local Green's function by solving the Dyson equation via
\begin{equation}
G^c_{loc}(t,t')=\sum_{\bf k}\left [ (G^c_{loc,non})^{-1}({\bf k})-\Sigma^c\right ]^{-1}(t,t')
\label{eq: local}
\end{equation}
where the Green's function and self-energy are continuous matrix operators in time, and we take the $(t,t')$ matrix elements after the inverse. This equation is solved by discretizing it on the contour and employing the method of solving the equation of motion differential equation via direct integration~\cite{dutch}. When properly formulated, the algorithm for doing this is highly efficient. Eq.~(\ref{eq: local})  includes the noninteracting nonequilibrium Green's function on the lattice, which can be found analytically~\cite{frtur_prb71}. Once we have the local Green's function, we are ready to solve the impurity problem. In this case, we do not need to determine the effective medium and solve the full impurity problem because the expression for the self-energy depends only on the local Green's function, hence we have
\begin{equation}
\Sigma^c(t,t')=ig^2 D^c(t,t')G^c_{loc}(t,t')
\label{eq: pert}
\end{equation}
which has to be solved self-consistently because the local Green's function depends on $\Sigma^c$, and the dressed phonon propagator depends on the Green's function. The dressed phonon propagator is constructed from the initial equilibrium propagator, which given by
\begin{equation}
D^c_0(t,t')=-i[n_B(\Omega)+1-\theta_c(t,t')]e^{i\Omega(t-t')}-i[n_B(\Omega)+\theta_c(t,t')]e^{-i\Omega(t-t')}.
\label{eq: phonon}
\end{equation}
Here, $n_B(\Omega)=1/[\exp(\beta\Omega)-1]$ is the Bose distribution function and $\theta_c(t,t')$ is equal to one if $t$ is ahead of $t'$ on the contour and is zero otherwise. 
We employ a conserving approximation where the electron and phonon self-energies are derived from a single functional.  This results, in addition to the Dyson equation for the electrons, in one for the phonon:
\begin{align}
D^{c}(t, t') = D_{0}^{c}(t,t ')+\int \int dt_1dt_2D_0^{c}(t,t_1)\Pi^{c}(t_1, t_2)D^{c}(t_2,t').
\end{align}
The phonon self-energy $\Pi^{c}(t,t')$ is obtained from the electron Green's functions through:
\begin{align}
\Pi^{c}(t, t') = -i G^{c}(t,t') G^{c}(t',t)
\end{align}
where the product is to be evaluated through the Langreth rules.
More details can be found elsewhere~\cite{michael_prx,lex_prb}.

To summarize how the Green's functions are calculated, we do the following: (i) we decide what the initial temperature $T$ is of the system before the field is turned on; (ii) we incorporate the field via a spatially uniform vector potential that is oriented along the diagonal direction; and (iii) we iterate the nonequilibrium Migdal-Eliashberg theory (or equivalently, the NEDMFT) until both self-energies converge. At this stage, we have both the contour-ordered self-energy and the contour-ordered Green's function for both the electrons and the phonons. Note that we work with the full local Green's function here, summing over all momenta in the Brillouin zone.

We extract the data in a similar fashion to what would be done in experiment. We first construct the time-resolved photoemission spectroscopy signal, neglecting matrix element effects~\cite{trpes}, which is formed from the lesser Green's function and the envelope of the probe pulse, $s(t)=\exp[-(t-t_0)^2/2\sigma^2_{pr}]/(\sqrt{2\pi}\sigma_{pr})$, centered at $t_0$, with a spread (standard deviation) given by $\sigma_{pr}$ (we use $\sigma_{pr}=25$~eV$^{-1}$ in this work). The formula is
\begin{equation}
P(\omega;t_0)\propto{\rm Im}\int dt \int dt' s(t)s(t')G^<_{loc}(t,t')e^{i\omega(t-t')}.
\label{eq: trpes}
\end{equation}
There is a small subtlety in what we do to determine the PES signal, and we do this to match results closer to how some experiments have been performed~\cite{lanzara}. Namely, we first construct the gauge-invariant Green's function and then we perform a partial summation over momentum, by summing only over the diagonal direction for {\bf k}. This is then used to approximate the TR-PES signal, which should be a good approximation because any anisotropy is washed out by the electron-phonon scattering, as in low-temperature superconductors.
After computing the TR-PES signal, the frequency axis is divided into contiguous bins (with a width of 0.01) and we integrate the total signal within each bin and plot as a function of the probe time $t_0$. These populations then decay as a function of time, with a typical exponential decay. To find the exponent, we either fit the tail of the curve to an exponentially decaying curve, or we extract the time-constant for the decay directly by numerically calculating the derivative and dividing the derivative by the function to give the decay rate (under the assumption that the system is decaying exponentially). As we will see below, both methods give similar results for the decay rate at a given time.

We end this section with a discussion about linear response. Within DMFT, the optical conductivity has no vertex corrections for the linear-response regime~\cite{no_vertex}. Hence, we can evaluate the dc conductivity in terms of a many-body transport relaxation time~\cite{freericks_book,arsenault}
\begin{equation}
\sigma_{dc}\propto \int d\omega \left (-\frac{df(\omega)}{d\omega}\right )\tau(\omega)
\label{eq: sigma_dc}
\end{equation}
where $f(\omega)=1/[1+\exp(\beta\omega)]$ is the Fermi-Dirac distribution function and in two-dimensions, we have
\begin{equation}
\tau(\omega)=\frac{1}{2\pi^2}\left [ \frac{{\rm Im}G^R_{trans}(\omega)}{{\rm Im}\Sigma^R(\omega)}+\frac{1}{8}-\frac{1}{8}{\rm Re}[\{w+\mu-\Sigma^R(\omega)\}G^R(\omega)]\right ],
\label{eq: relax_time}
\end{equation}
where the $R$ superscript means retarded, and the transport Green's function is defined with an extra $v^2$ (square of the band velocity) in its definition, so that
\begin{equation}
G^R_{trans}(\omega)=\sum_{\bf k}v^2({\bf k})\frac{1}{\omega-\epsilon({\bf k})-\Sigma^R(\omega)}.
\label{eq: g_trans}
\end{equation}
Here $v=\nabla_{\bf k}\epsilon({\bf k})$ is the band velocity.

Unfortunately, this formula does not easily generalize to the nonequilibrium limit. But in the spirit of this result, we examine a relaxation time similarly constructed from the retarded transport Green's function and the retarded self-energy at the same average time; we neglect the second and third terms in Eq.~(\ref{eq: relax_time}) since this is likely to only be semiquantitative result at best, and we assume the derivative of the Fermi-Dirac distribution is so sharp it can be approximated by a delta function for each population at a given energy above the chemical potential.
Our ansatz is that we compare
\begin{equation}
\bar\tau(\omega,t_{ave})=\frac{1}{\eta(\omega,t_{ave})}\propto \frac{{\rm Im}G^R_{trans}(\omega,t_{ave})}{{\rm Im}\Sigma^R(\omega,t_{ave})}
\label{eq: taubar}
\end{equation}
against the calculated decay rates extracted from analyzing the photoemission spectra, with an overall normalization factor still to be determined. We call this transport-based relaxation rate $\eta$.

\begin{figure}[htb]
 \centerline{\includegraphics[width=0.5\textwidth]{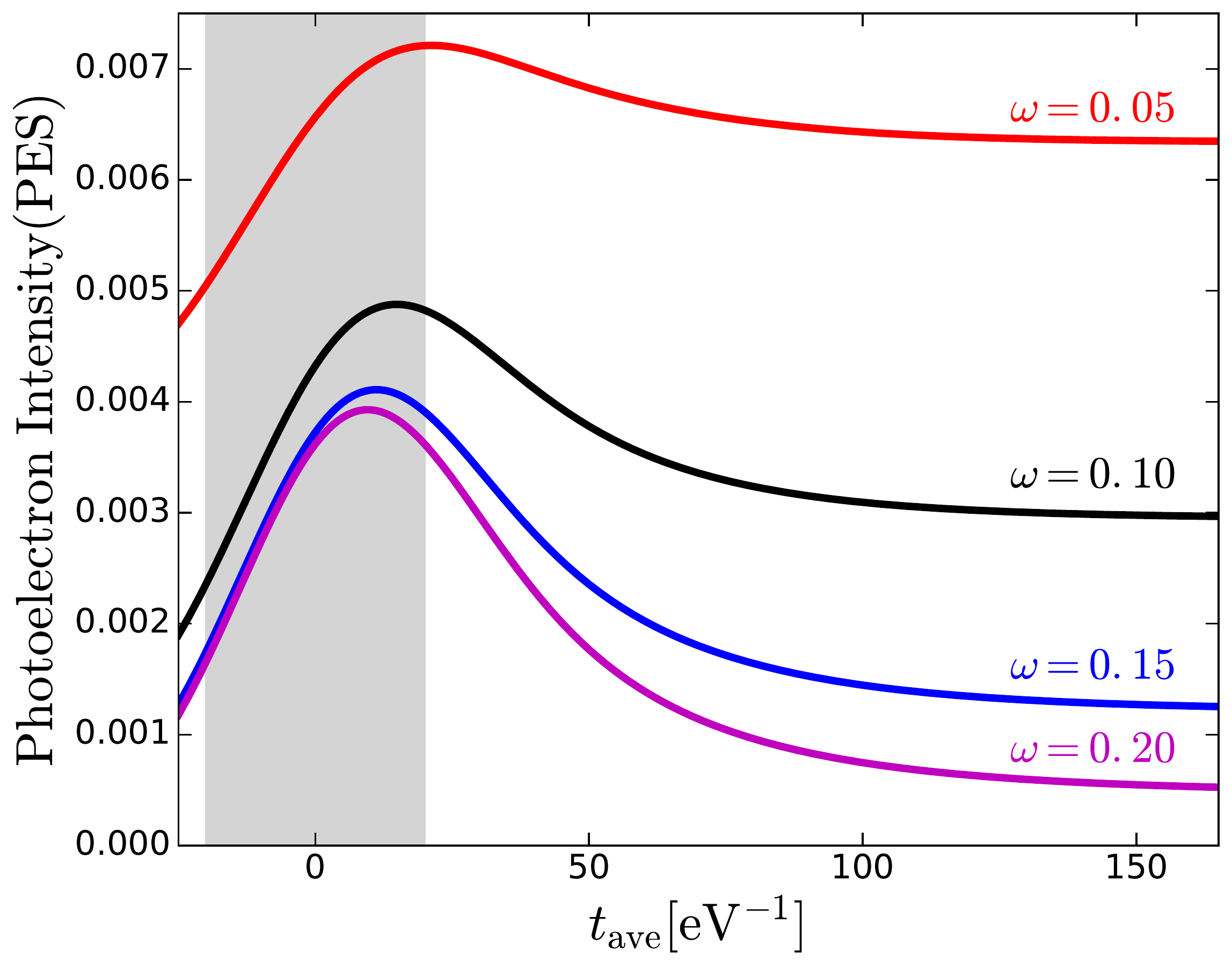}}
 \caption{(Color online.) Time-resolved photoemission spectra for different frequency bins as a function of time. One can see that the higher frequencies decay much faster than the lower frequencies due to the phonon window effect  (here, $\Omega=0.1$~eV). The grayed region is where the field is on, which has a total width of about 50~eV$^{-1}$.}
 \label{fig: pes}
\end{figure}

\section{Discussion and Results}
\label{sec:3}

\begin{figure}[htb]
 \centerline{\includegraphics[width=0.5\textwidth]{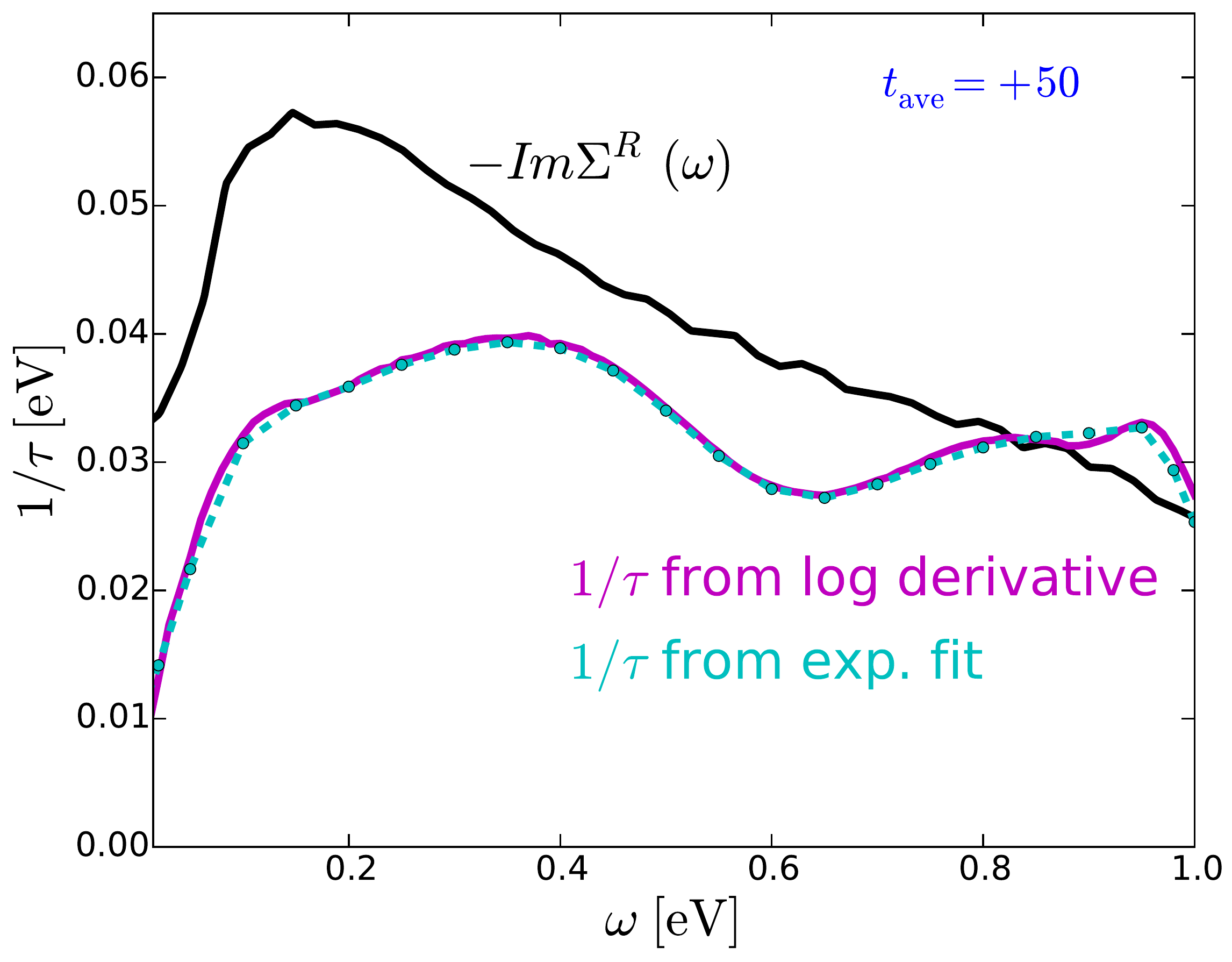}}
 \caption{(Color online.) Relaxation rate for different frequency bins at $t_{ave}=50$~eV$^{-1}$. The imaginary part of the retarded equilibrium self-energy is shown for comparison. While there is semiquantitative agreement here, they clearly differ. }
 \label{fig: relax_rate}
\end{figure}

In Fig.~\ref{fig: pes}, we plot the time-resolved photoemission spectra for different frequency bins as a function of time. The grayed region is the region where the pump is on. We take the data for different frequencies and times and extract an effective relaxation time in one of two different ways. The first way is to fit the data according to an exponential fit for the data in a time window about the given time. The second method is to extract an effective exponential relaxation time by computing the numerical derivative and dividing by the function at a given time; that is, by computing the logarithmic derivative. This satisfies
\begin{equation}
\frac{1}{\tau(\omega,t_{ave})}=-\frac{dP(\omega,t_{ave})}{dt_{ave}}\frac{1}{P(\omega,t_{ave})}.
\label{eq: log_deriv}
\end{equation}
The results are plotted in Fig.~\ref{fig: relax_rate}. One can see that both methods for extracting the instantaneous decay rate agree to high accuracy, but the overall relaxation rate differs from the equilibrium self-energy result.

\begin{figure}[htb]
 \centerline{\includegraphics[width=0.48\textwidth]{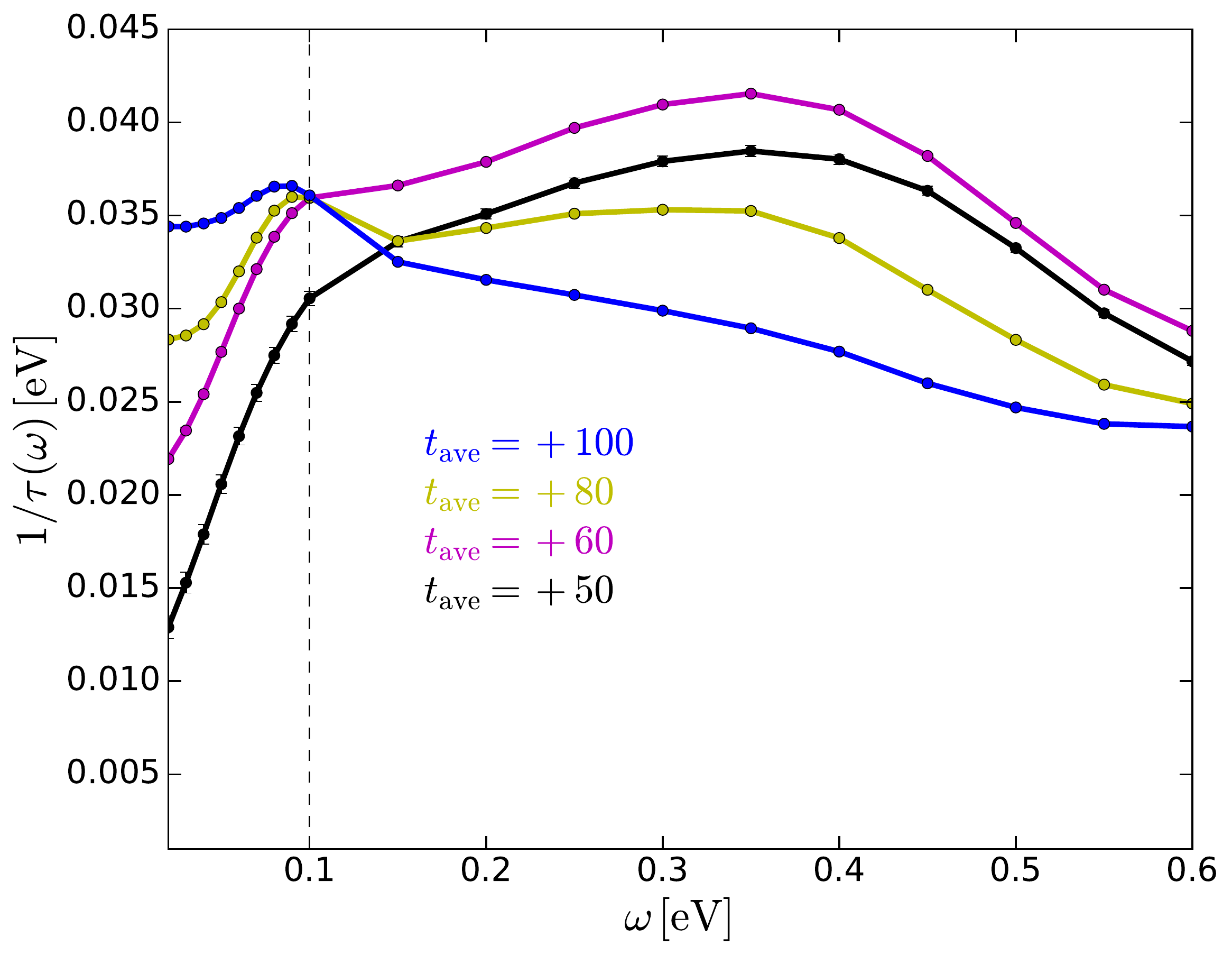}~~\includegraphics[width=0.48\textwidth]{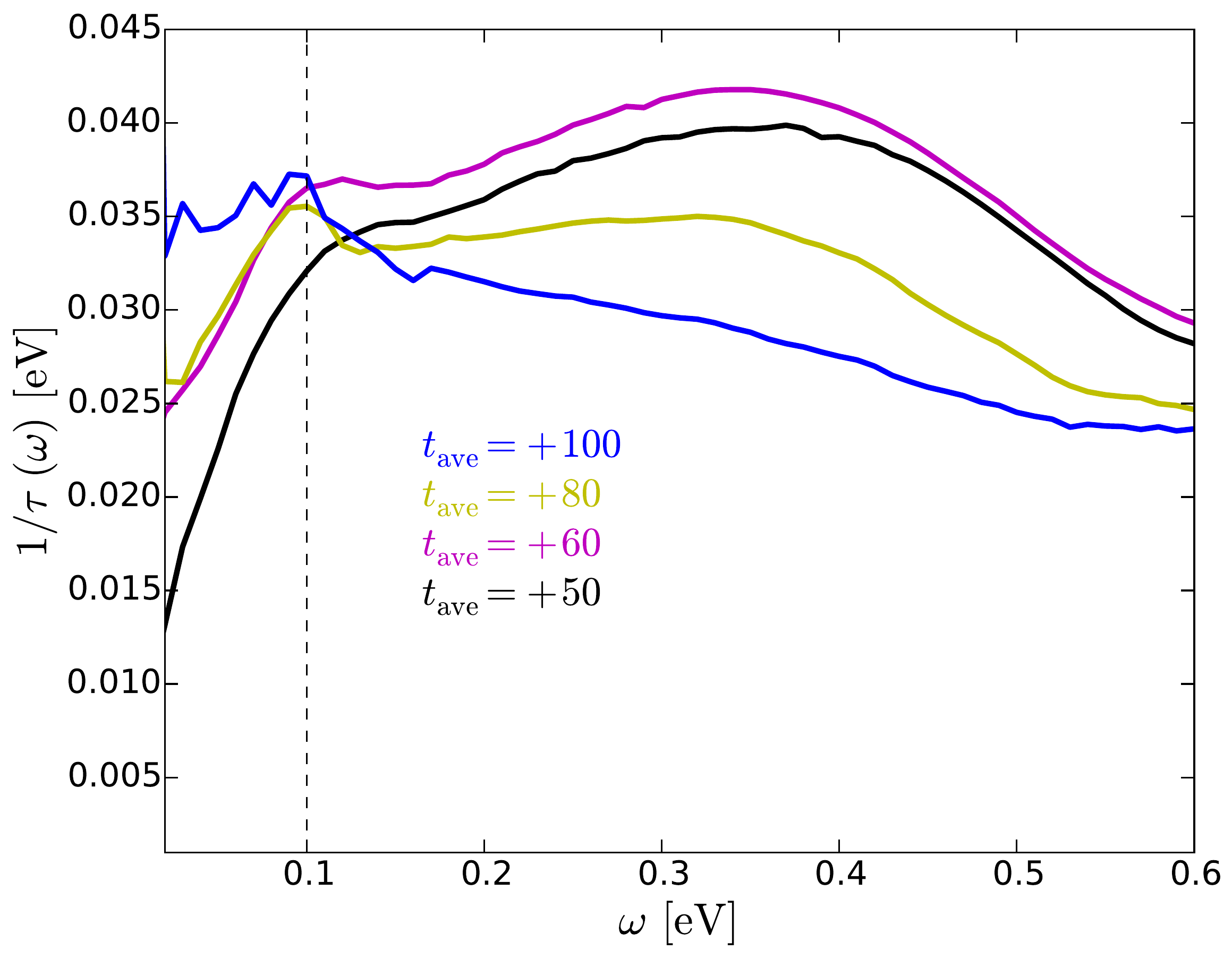}}
 \caption{(Color online.) Left: relaxation rate as determined from a fit to an exponential decay for different average times. The phonon frequency is marked by a dashed line to help identify the phonon window effect. Right: a similar plot for the relaxation rates as extracted from a logarithmic derivative.}
 \label{fig: relax_rate2}
\end{figure}

We next compare these relaxation rates, extracted with the two different methods, for different average times in Fig.~\ref{fig: relax_rate2}. We can see a phonon window effect for short times, which disappears at longer times, and we can see a generic behavior that produces a fairly flat response for the relaxation rate over these frequency values.

\begin{figure}[htb]
\centerline{ \includegraphics[width=0.5\textwidth]{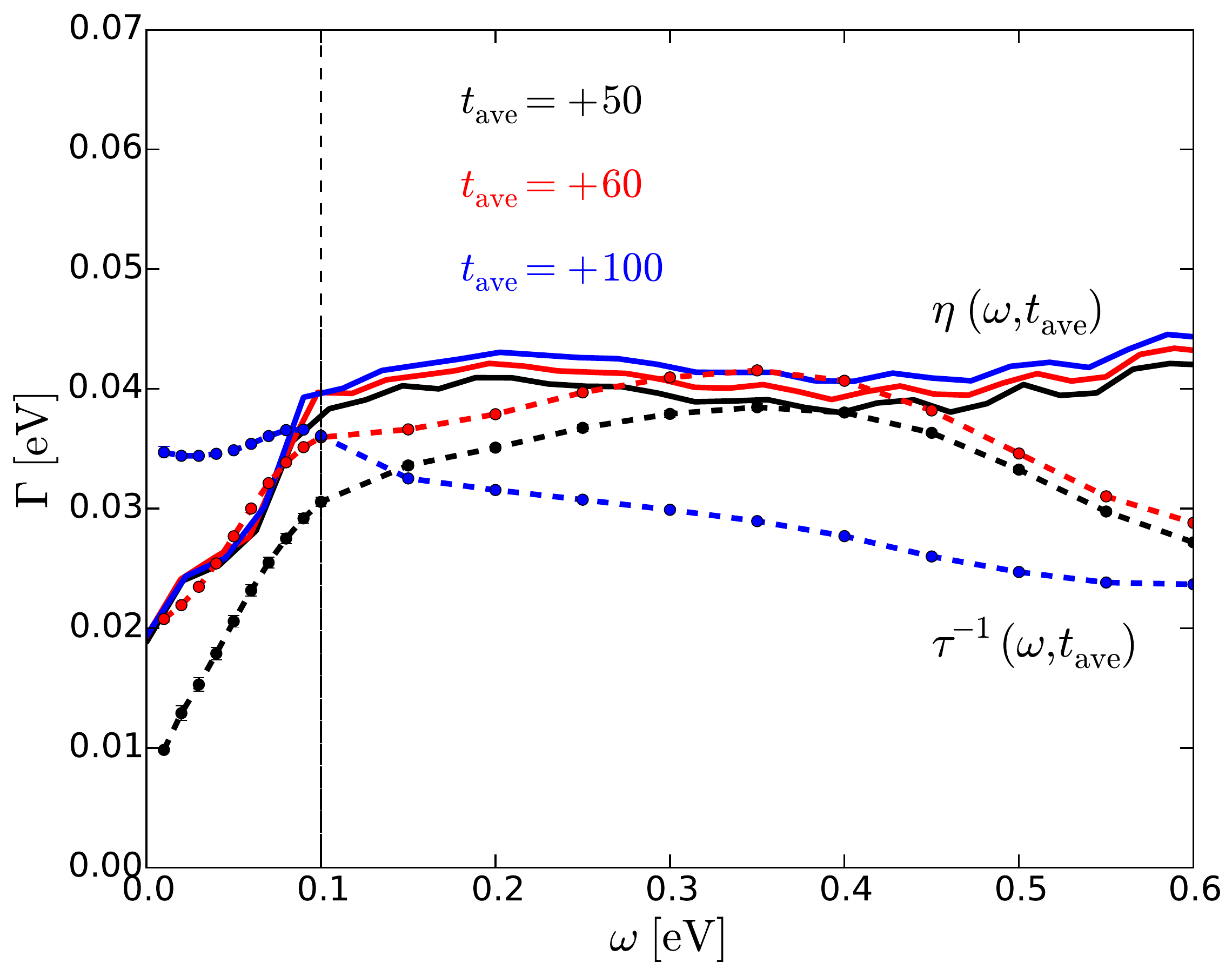}}
 \caption{(Color online.) Transport-based relaxation rate (renormalized for best fit) compared to the relaxation rate extracted from the TR-PES data. One can see it is much flatter outside the phonon window, in good agreement with the data, but it does not show a disappearance of the phonon window effect at long times like the data do (compare the blue lines for the longest times).}
\label{fig: eta}
\end{figure}

Finally, we form the result from the transport Green's function and compare with the relaxation rate data in Fig.~\ref{fig: eta}. The overall factor is about 6. One can immediately conclude that this transport Green's function-based relaxation time is much flatter in frequency, in agreement with the data, but it does not show the strong average time-dependence within the phonon window, where the phonon window effect disappears for long average times.

\section{Conclusions}
\label{sec:4}

Nonequilibrium relaxation is complicated. Indeed, the simple notions for how to determine it from the self-energy are known to fail when the electron correlations become strong. Here, we have made an initial attempt to remedy this problem by considering the modification of the relaxation rate due to a generalization of the transport relaxation time to nonequilibrium. We find that while this approach does do better in modeling the weak frequency dependence of the data outside the phonon window, it does not properly show the evolution of this dependence inside the window, especially for long times. So, this might be a step in the right direction, but, it unfortunately has an adjustable parameter for the overall relaxation rate, which needs to be determined for this system, and which provides less predictive power than if we had a prediction on an absolute scale.

In the future, we hope to be able to find an even better ansatz for the nonequilibrium relaxation time in order to find the microscopic origin of this relaxation. The nonequilibrium behavior is far more complex than the linear-response regime and we need more work to both have good data to compare with and to determine the proper microscopic basis for the relaxation, including all bottleneck effects.

\begin{acknowledgments}
This work was supported by the Department of Energy, Office of Basic Energy Sciences, Division of Materials Sciences and Engineering under Contract No.  DE-FG02-08ER46542 (Georgetown). Computational resources were provided by the National Energy Research Scientific 
Computing Center supported by the Department of Energy, Office of Science, under Contract No. DE- AC02-05CH11231. J.K.F. was also supported by the McDevitt Bequest at Georgetown. 
\end{acknowledgments}

\end{document}